\documentstyle[aps,preprint,eqsecnum]{revtex}
\def\footnoterule{\kern-3pt \hrule width \hsize \kern2.6pt}

\begin{document}
\draft
\tighten
\preprint{CTP\#2243}

\title{
Scalar Field Quantization\\
on the 2+1 Dimensional Black Hole Background\cite{doe}}

\author{Gilad Lifschytz and Miguel Ortiz}

\address{{~} \\
Center for Theoretical Physics \\
Laboratory for Nuclear Science and \\
Department of Physics \\
Massachusetts Institute of Technology \\
Cambridge, MA~~02139, USA \\
e-mails: gil1 and ortiz @mitlns.mit.edu
{~}}
\vskip 1 true cm
\date{Submitted to Physical Review D}

\maketitle
\begin{abstract}
The quantization of a massless conformally coupled scalar field on
the 2+1 dimensional Anti de Sitter black hole background is presented.
The Green's function is calculated, using the fact that the black hole
is Anti de Sitter space with points identified, and taking into account the
fact that the black hole
spacetime is not globally hyperbolic.
It is shown that the Green's function calculated in this way
is the Hartle-Hawking Green's function. The Green's function is used to
compute $\langle T^\mu_\nu \rangle$, which  is regular
on the black hole horizon, and diverges at the singularity.
A particle detector response function outside the
horizon is also calculated and shown to
be a fermi type distribution.
The back-reaction from $\langle T_{\mu\nu} \rangle$ is calculated
exactly and is shown to give rise to a curvature
singularity at $r=0$ and to shift the horizon outwards.
For $M=0$ a horizon develops, shielding the singularity. Some
speculations about the endpoint of evaporation are discussed.\\
\end{abstract}

\section*{Introduction}

The study of black hole physics is complicated by the many technical and
conceptual problems associated with quantum field theory in curved
spacetime. One serious difficulty
is that exact calculations are almost impossible
in 3+1 dimensions. In this paper we shall instead study
some aspects of quantum field theory on a 2+1 dimensional
black hole background. This enables us to
obtain an exact expression for the Green's
function of a massless, conformally coupled scalar field
in the Hartle-Hawking vacuum \cite{HnH}. We use this Green's
function to study particle creation by the black
hole, back-reaction and the endpoint of evaporation.

We shall work with the 2+1 dimensional black hole solution found
by Ba\~nados, Teitelboim and Zanelli (BTZ) \cite{BTZ}.
It had long been thought that black holes cannot exist in
2+1 dimensions for the simple reason that there is no gravitational
attraction, and therefore no mechanism for confining large densities of
matter. This difficulty has been circumvented in the BTZ
spacetime\footnote{A charged black hole solution in 2+1 dimensions had
previously been found in Ref.~\cite{Reznik}. For further discussions on the
BTZ black hole, see Refs. \cite{others}.}, but not
surprisingly, their solution has some features that we do not normally
associate with black holes in other dimensions, such as the absence of a
curvature singularity. It is interesting to ask
whether this spacetime behaves quantum mechanically in a way
consistent with
more familiar back holes.

The spinless BTZ spacetime has a metric
\cite{BTZ}
\[
ds^2 = -N^2 dt^2 + N^{-2} dr^2 + r^2 d\phi^2
\]
where
\[
N^2 = \frac{r^2 - r_+^{\,2}}{\ell^2},\qquad r_+^{\,2}=M\ell^2.
\]
Here $M$ is the mass of the black hole.
The metric is a solution to Einstein's equations with a negative
cosmological constant, $\Lambda=-\ell^{-2}$, and
the curvature of the black hole solution is constant everywhere.
As a result there is no curvature singularity as $r\to 0$.
A Penrose diagram of the spacetime is given in Fig.~\ref{fig:penrosebh}.

An important feature of the BTZ solutions is that the solution
with $M=0$ (which we refer to as the vacuum solution),
is not AdS$_3$. Rather, it is a solution that is not
globally Anti de Sitter invariant. It has no horizon, but does have
an infinitely long throat for small $r>0$, which is reminiscent of the
extreme Reissner-Nordstr\"om solution in 3+1 dimensions. It is worth
noting that there are other similarities between the spinless BTZ black
holes, $M\ge 0$, and the Reissner-Nordstr\"om solutions for $M\ge Q$. In
particular, the temperature associated with the Euclidean continuation of
the BTZ black holes has been computed in \cite{BTZ}, and it was found to
increase with the mass, and to decrease to zero as
$M\to 0$.
Thus, if we carry over the usual notions from four dimensional black
holes, the $M=0$ solution appears to be a stable endpoint of evaporation.

A feature of the BTZ solution that we shall make use of, is that
the solution arises from identifying points in
AdS$_3$, using the orbits of a spacelike Killing vector field. It is this
property that is the starting point of our derivation of a Green's function
on the black hole spacetime. We construct a Green's function on AdS$_3$,
and this translates to a Green's function on the black hole via the method
of images.

A Green's function constructed in this way is only interesting if we can
identify the vacuum with respect to which it is defined. We prove that
our construction gives the Hartle-Hawking Green's function.
It is worth noting that for the BTZ black
hole, there is a limited choice of vacua.
Quantisation on AdS$_3$ is hampered by the fact that AdS$_3$ is not
globally hyperbolic, and this necessitates the use of boundary conditions
at spatial infinity \cite{Isham} (see Fig.~\ref{fig:penroseads}), as
discussed in Appendix A.
This problem carries
over to the black hole solution, and as a result, the
value of the field at spatial infinity is governed by either Dirichlet or
Neumann type boundary conditions. Thus a Cauchy surface for the region R of
the BTZ black hole is either the past or the future horizon only. With this
knowledge, the  natural definition of the Hartle-Hawking vacuum is with
respect to Kruskal modes on either horizon, whereas there does not appear
to be a natural definition of an Unruh vacuum (see \cite{fullig,GibbnPerr}
for a
discussion of the various eternal black hole vacua). The definition of an
Unruh vacuum might be possible given a description of the formation
of a BTZ black hole from the vacuum via some sort of infalling matter,
but as far as we are aware, no such construction has been found.

Having an explicit expression for the Hartle-Hawking Green's function,
we are able to obtain a number of exact results.
As a check, we show that it satisfies the
KMS thermality condition \cite{KMS}.
We then compute the expectation value of the
energy-momentum tensor and the response of a particle detector for both
nonzero $M$ and for the vacuum solution. For nonzero $M$ we
address the issue of whether the response of the particle detector can be
interpreted as radiation emitted from the black hole, although a clear
picture does not emerge.

For the $M=0$ solution, we find a non-zero energy-momentum tensor,
although the corresponding Green's function is at zero temperature,
and there is no particle detector response. We
interpret this as a sort of Casimir energy.
Classically, the vacuum solution appears to be similar to the extremal
Reissner-Nordstr\"om solution, in the sense that we expect that
if any matter is thrown in,
a horizon develops. Quantum mechanically, the $M=0$ solution appears to be
unstable to the formation of a horizon, when the back-reaction caused by
the Casimir energy is taken into account. This suggests that
the endpoint of evaporation may not look like the classical $M=0$ solution.

The paper is organized as follows.
In section~\ref{sec:i} we study the 1+1 dimensional solution which arises
from a dimensional reduction of the BTZ black hole \cite{ortiz},
and show that the vacuum defined by
the Anti de Sitter (AdS) modes is the same as that defined by
the Kruskal modes; with this encouraging result we tackle the 2+1 case.
Section~\ref{sec:ii} contains a review
of the essential features of the geometry of the BTZ black hole.
In section~\ref{sec:iii} we construct the Wightman Green's functions
on the black hole spacetime from the AdS$_3$ Wightman function, using the
method of images.  We then show
that the Green's function coincides with the Hartle-Hawking
Green's function \cite{HnH},
by showing that it is analytic
and bounded in the lower half of the complex
$\bar{V}$ plane on the past horizon $(\bar{U}=0)$, where $\bar{V}, \bar{U}$
are the Kruskal null coordinates. We also compute the Wightman function for
the $M=0$ solution, and compare this to the $M\to 0$ limit of the results
for $M\ne 0$. Section \ref{sec:iv} contains a calculation of
$\langle T_{\mu\nu} \rangle$ for all $M\ge 0$. For the black hole
solutions, it is regular on the horizon, and for all $M$ it is
singular as $r\to0$. In Section \ref{sec:v}
the response function of a stationary particle detector outside the horizon
is calculated and shown to be of a fermi type distribution. A discussion is
given of how this response might be interpreted.
In Section \ref{sec:vi} we calculate the back reaction
induced by $\langle T_{\mu\nu}
\rangle$, and show that the spacetime develops a
curvature singularity and a larger horizon for a given $M$.
Throughout this paper we use metric signature $(-++)$, and natural units in
which $8G=\hbar=c=1$.

\section{2-D Black Hole}
\label{sec:i}

Let us begin by looking at quantum field theory on the region of
Anti de Sitter
spacetime in 1+1 dimensions described by the metric
\[
ds^2 = - \left( \frac{r^2 - M \ell^2}{\ell^2} \right) dt^2
+ {\left( \frac{r^2 - M \ell^2}{\ell^2} \right)}^{-1}dr^2
\qquad 0 < r < \infty \qquad-\infty < t < \infty,
\]
where $M$ is the mass of the solution.
This metric was discussed in \cite{ortiz} as the dimensional reduction of
the spinless BTZ black hole, and can be thought of as being a region of
AdS$_2$ in Rindler-type co-ordinates.
Under the change of co-ordinates
\[
r = \sqrt{M\ell^2} \sec \rho \cos\lambda,
\qquad\tanh \left( \frac{\sqrt{M} t}{\ell} \right)
= \frac{\sin \rho}{\sin\lambda}
,
\]
where we shall call $(\lambda,\rho)$ AdS co-ordinates,
the metric becomes
\[
ds^2 = \ell^2 \sec^2\rho (-d\lambda^2 + d\rho^2)
\]
which for
$-\frac{\pi}{2} \leq \rho \leq \frac{\pi}{2}$ and
$-\infty < \lambda < \infty$
is just AdS$_2$ \cite{HandE}.

It is possible to define Kruskal-like co-ordinates for this black hole,
which do not coincide with the usual AdS co-ordinates. For $r>M\ell^2$,
they are:
\begin{eqnarray}
U &=& \left( \frac{r - \sqrt{M} \ell}{r + \sqrt{M} \ell} \right)^\frac{1}{2}
\cosh \frac{\sqrt{M}}{\ell} t
\label{1.4}
\\
V &=& \left( \frac{r - \sqrt{M} \ell}{r + \sqrt{M} \ell} \right)^\frac{1}{2}
\sinh \frac{\sqrt{M}}{\ell} t.
\label{1.5}
\end{eqnarray}
The metric then takes the form
\[
ds^2 = \frac{- 2 \ell^2}{1 + \bar{U}\bar{V}} \, d\bar{U} d\bar{V}
\]
where $\bar{U}=V+U$, $\bar{V} = V-U$,
and the transformation between Kruskal and AdS
co-ordinates is
\[
\bar{U} = \tan \left( \frac{\rho + \lambda}{2} \right)\qquad
\bar{V} = \tan \left( \frac{\rho - \lambda}{2} \right) \,
\]
which is valid over all the Kruskal manifold.
The Kruskal co-ordinates cover only the part of AdS$_2$ with
\[
-\frac{\pi}{2} \leq \rho \leq \frac{\pi}{2} \qquad
-\frac{\pi}{2} < \lambda < \frac{\pi}{2}.
\]
We shall now show that the notion of positive frequency
in $(\lambda,\rho)$ (AdS)
modes coincides with that defined in $(U,V)$ (Kruskal) modes.

The AdS modes for a conformally coupled scalar field are normalized
solutions of
$\Box\psi = 0$, subject to the boundary conditions
\[
\phi \left( \rho = \frac{\pi}{2} \right)
= \phi \left( \rho = \frac{-\pi}{2} \right) = 0.
\]
The positive frequency modes are then
\begin{eqnarray}
\phi_m &=& \frac{1}{\sqrt{m\pi}} e^{-im\lambda} \sin m \rho
\qquad m {\rm ~even~} \geq 0 \nonumber
\\
\phi_m &=& \frac{1}{\sqrt{m\pi}} e^{-im\lambda} \cos m \rho
\qquad m {\rm ~odd~} \geq 0
\nonumber
\end{eqnarray}
and these define a vacuum state $|0\rangle_A$ in the usual way.

The Kruskal modes are solutions of
$\Box \psi = 0$
with  the boundary condition
$\psi(\bar{U}\bar{V} = -1) = 0$.
Positive frequency solutions are given by
\[
\psi_\omega = N_\omega\left(e^{-i\omega\bar{U}} - e^{i\omega/\bar{V}}
\right)\qquad \omega > 0
\]
where $N_\omega=(8\pi\omega)^{-1/2}$, and these define $|0\rangle_K$.
These modes are analytic and bounded in the lower half of the complex
$\bar{U}, \bar{V}$ plane.  In order to show
equivalence of the two vacua
$|0\rangle_A$ and $|0\rangle_K$,
it is enough to show that the positive frequency
AdS modes can be written as a sum of only positive frequency Kruskal modes.
Because of the analyticity properties of the Kruskal modes, it is enough to
show that the AdS modes are analytic and bounded in the lower half of the
complex $\bar{U}, \bar{V}$ plane \cite{fullig,BnD}.
Changing co-ordinates, we have
\begin{eqnarray}
\phi_m &=& \frac{1}{\sqrt{\pi} m 2 i}
\left( e^{-2im\arctan\bar{V}} - e^{-2im\arctan \bar{U}} \right)
\qquad m {\rm ~even}
\label{1.14}
\\
\phi_m &=& \frac{1}{\sqrt{\pi} m 2 i}
\left( e^{-2im\arctan\bar{V}} + e^{-2im\arctan \bar{U}} \right)
\qquad m {\rm ~odd.}
\label{1.15}
\end{eqnarray}
Using the definition $\arctan z = \frac{1}{2i} \ln \frac{1+iz}{1-iz}$
\cite{HandR}, (\ref{1.14}) and (\ref{1.15}) become
\[
\phi_m = \frac{1}{2\sqrt{\pi} m i}
\left[
\left(\frac{1-i\bar{V}}{1+i\bar{V}}\right)^m
\mp
\left(\frac{1-i\bar{U}}{1+i\bar{U}}\right)^m
\right]
\]
where $\pm$ is for $m$ odd or even.
These modes
can easily be seen to be
bounded and analytic in the lower half of the complex
$\bar{U}, \bar{V}$ plane for all $m$.
This establishes that the vacuum defined by the AdS modes is the same as
that defined by the Kruskal modes. Thus a Green's function defined on this
spacetime using AdS co-ordinates $(\lambda,\rho)$
corresponds to a Hartle-Hawking
Green's function, in the sense discussed in the Introduction.

\section{The Geometry of the 2+1 Dimensional Black Hole}
\label{sec:ii}

In this paper, we shall be working only with the spinless
black hole solution in 2+1
dimensions
\begin{equation}
ds^2 = -N^2 dt^2 + N^{-2} dr^2 + r^2 d\phi^2
\label{2.1}
\end{equation}
where
\[
N^2 = \frac{r^2 - r_+^{\,2}}{\ell^2},\qquad r_+^{\,2}=M\ell^2.
\]

As was shown in \cite{BTZ}, this metric has constant curvature,
and is a portion of three dimensional Anti de Sitter space with
points identified. The identification is made using
a particular
killing vector $\xi$, by identifying all points $x_n = e^{2\pi n i \xi} x$.
In order to see this most clearly, it is useful to
introduce different sets of co-ordinates on AdS$_3$.

AdS$_3$ can be defined as the surface $-v^2-u^2+x^2+y^2=-\ell^2$ embedded in
$R^4$ with metric $ds^2 = -du^2 - dv^2 + dx^2 + dy^2$.
A co-ordinate
system $(\lambda,\rho,\theta)$
which covers this space, and which we shall refer to as AdS
co-ordinates,  is defined by \cite{Isham}
\begin{eqnarray}
u &=& \ell \cos \lambda \sec \rho\qquad
v = \ell \sin \lambda \sec \rho\nonumber\\
x &=& \ell \tan \rho \cos \theta \qquad
y = \ell \tan \rho \sin \theta
\nonumber
\end{eqnarray}
where
$0 \leq \rho \leq \frac{\pi}{2}$, $  0 < \theta \leq 2\pi$, and
$0 < \lambda < 2\pi$.
In these co-ordinates, the AdS$_3$ metric becomes
\[
ds^2 =
\ell^2 \sec^2 \rho (- d\lambda^2 + d\rho^2 + \sin^2 \rho \, d\theta^2 ).
\]
AdS$_3$ has topology $S^1$ (time) $\times R^2$ (space) and hence
contains closed timelike curves.  The angle $\lambda$ can be unwrapped to form
the covering space of AdS$_3$, with $-\infty<\lambda<\infty$, which does not
contain any closed timelike curves. Throughout this
paper we work with this covering space, and this is what we henceforth
refer to as AdS$_3$. As mentioned in the Introduction, even this covering
space presents difficulties since it is not globally hyperbolic (see the
discussion in Appendix A).

The identification taking AdS$_3$ into the
black hole (\ref{2.1}) is most easily
expressed in terms of co-ordinates $(t,r,\phi)$, related in an obvious way
to those used above, and defined on AdS$_3$ by
\[
\begin{array}{l}
u = \sqrt{A(r)} \cosh \left(\frac{r_+}{\ell} \phi\right)
\\
x = \sqrt{A(r)} \sinh \left(\frac{r_+}{\ell} \phi \right)
\\
y = \sqrt{B(r)} \cosh \left(\frac{r_+}{\ell^2} t \right)
\\
v = \sqrt{B(r)} \cosh \left(\frac{r_+}{\ell^2} t\right)
\end{array}
\qquad\qquad r > r_+
\]
\[
\begin{array}{lr}
u = \sqrt{A(r)} \cosh \left(\frac{r_+}{\ell} \phi \right)
\\
x = \sqrt{A(r)} \sinh \left(\frac{r_+}{\ell} \phi\right)
\\
y = -\sqrt{-B(r)} \cosh \left(\frac{r_+}{\ell^2} t \right)
\\
v = \sqrt{-B(r)} \cosh \left(\frac{r_+}{\ell^2} t\right)
\end{array}
\qquad\qquad 0 > r > r_+.
\]
Note that $-\infty<\phi<\infty$. Under the identification $\phi \to \phi +
2\pi n$, where $n$ is an integer, these regions of AdS$_3$ become regions
R and F of the black hole. Regions P and L are defined in an analogous way
\cite{BTZ} (see Fig.~\ref{fig:penrosebh} for a definition of regions
F (future), P (past), R (right) and L (left)).
The $r=0$ line is a line of fixed points under this identification, and
hence there is a singularity there in the black hole spacetime of the
Taub-NUT type \cite{BTZ,HandE}.

Finally, it is possible to define Kruskal co-ordinates on the black hole.
The relation
between the Kruskal co-ordinates $V$ and $U$ and the black hole
co-ordinates $t$ and $r$ is precisely as in (\ref{1.4}) and (\ref{1.5}).
$U,V$ and an
unbounded $\phi$ cover the region of AdS$_3$ which becomes the black hole
after the identification.

\section{Green's Functions on the 2+1 Dimensional Black Hole}
\label{sec:iii}

In this section we derive a Green's function on the black hole spacetime, by
using the method of images on a  Green's function on AdS$_3$.
We then show that the resulting Green's function is thermal, in that it
obeys a KMS condition \cite{KMS}. Using the
analyticity properties discussed in the Introduction, the Green's function
is also shown to be defined with respect to
a vacuum state corresponding to Kruskal
co-ordinates on both the past and future horizons of the black hole.
We therefore interpret it as a Hartle-Hawking
Green's function.
Finally we derive the Green's function for the $M=0$
solution directly from a mode sum, and compare
it with the $M\to 0$ limit of the black hole Green's function.

\subsection{Deriving the Green's Functions}
\label{sec:iii.1}

Since the black hole spacetime is given by identifying points on AdS$_3$
using a spacelike Killing vector field, we can use the method of images to
derive the
two point function on the black hole spacetime. Given the two point function
$G^+_A(x,x')$ on AdS$_3$,
\[
G^{\,+}_{\rm BH} (x, x';\delta) = \sum_n e^{-i\delta n} G^{+}_{\rm A} (x,
x'_n)
\]
Here $x_n'$ are the images of $x'$ and $0\le\delta<\pi$
can be chosen arbitrarily.  For a general $\delta$ the modes of
the scalar field on the black hole background will satisfy
$\phi_m(e^{2\pi n \xi} x) = e^{-i\delta n} \phi_m(x)$. $\delta=0$ for
normal scalar fields and $\delta=\pi$ for twisted fields.
{}From now on we will restrict ourselves to $\delta=0$.

This definition of the Green's function on the black hole spacetime
means that when
summing over paths to compute the Feynman Green's function $G_F(x,x')$, we sum
over all paths in AdS$_3$. Hence paths that cross and recross the singularities
must be taken into account (compare this with the results of Hartle and
Hawking \cite{HnH}).

As explained in Appendix A, boundary conditions at infinity must be imposed
on any Green's function on AdS$_3$ in order to deal with the fact that
AdS$_3$ is not globally hyperbolic. From Appendix A,
we have
\begin{equation}
G_A^{+} = G_{A1}^{+} \pm G_{A2}^{+}
\label{2.5}
\end{equation}
where $+(-)$ corresponds to Neumann (Dirichlet) boundary conditions (from
now on, it should be assumed that the upper (lower) sign is always
for Neumann (Dirichlet) boundary conditions unless otherwise stated).
The individual terms in (\ref{2.5}) are given by
\begin{eqnarray}
G^{+}_{A1} (x,x') &=& \frac{1}{4\sqrt{2} \pi \ell}
\left(\cos
(\Delta \lambda - i \epsilon) \sec \rho' \sec \rho - 1 - \tan \rho \tan
\rho' \cos \Delta\theta\right)^{-\frac{1}{2}} \nonumber
\\
G^{+}_{A2} (x,x') &=& \frac{1}{4\sqrt{2} \pi \ell}
\left(\cos
(\Delta \lambda - i \epsilon) \sec \rho' \sec \rho + 1 - \tan \rho \tan
\rho' \cos \Delta\theta\right)^{-\frac{1}{2}}.
\nonumber
\end{eqnarray}
$\Delta\lambda$ is defined as $\lambda-\lambda'$, and similarly for all
other co-ordinates.

The sign of the $i\epsilon$ is proportional to ${\rm sign}
(\sin\Delta\lambda)$. It
is only important for timelike separated points, for which the argument of
the square root is negative. In the three
dimensional Kruskal co-ordinates on AdS$_3$,
the identification is only in the angular direction.
For timelike separated points,
${\rm sign}\Delta\lambda={\rm sign}\Delta V$, where $V$ is the Kruskal time. It
follows that for all identified points the sign of $i\epsilon$ in
$G(x,x'_n)$ is the same.

We now work in the black hole
co-ordinates $(t,r,\phi)$, so that
the identification taking AdS$_3$ into the black hole spacetime
is given by $\phi \to \phi + 2 \pi n$.
Under this identification,
the two point function on the black hole background becomes
\[
G^{+}(x,x') = \frac{1}{4\sqrt{2}\pi\ell}
\left[ G_1^+ (x,x') \pm G_2^+ (x,x') \right]
\]
where for $x, x' \in $ region R
\begin{eqnarray}
G_1^{+} (x,x') &=& \sum_{n=-\infty}^\infty \left[
\frac{r r'}{r_+^2} \cosh\frac{r_+(\Delta \phi + 2 \pi n)}{\ell}
- 1 - \frac{(r^2-r_+^2)^{1\over2} (r'^2-r^2_+)^{1\over2}}{r_+^2}
\cosh\frac{r_+(\Delta t-i\epsilon)}{\ell^2}
\right]^{-\frac{1}{2}}
\label{3.13}
\\
G_2^{+} (x,x') &=& \sum_{n=-\infty}^\infty \left[
\frac{r r'}{r_+^2} \cosh\frac{r_+(\Delta \phi + 2 \pi n)}{\ell}
+ 1 - \frac{(r^2-r_+^2)^{1\over2} (r'^2-r^2_+)^{1\over2}}{r_+^2}
\cosh\frac{r_+(\Delta t-i\epsilon)}{\ell^2}
\right]^{-\frac{1}{2}}.
\label{3.14}
\end{eqnarray}
For $x,x' \in$ region F,
\begin{eqnarray}
G_1^+(x,x') &=& \sum_{n=-\infty}^\infty
\left[ \frac{r r'}{r_+^2} \cosh\frac{r_+(\Delta \phi + 2 \pi n)}{\ell}
- 1 +
 \frac{(r_+^2 - r^2)^{1\over2} (r_+^2 - r'^2)^{1\over2}}{r_+^2}
\cosh\frac{r_+\Delta t}{\ell^2}
 + i \epsilon {\rm ~sign~}\Delta V
\right]^{-{1\over2}} \nonumber\\
G_2^+(x,x') &=& \sum_{n=-\infty}^\infty
\left[ \frac{r r'}{r_+^2} \cosh\frac{r_+(\Delta \phi + 2 \pi n)}{\ell}
+ 1 +
 \frac{(r_+^2 - r^2)^{1\over2} (r_+^2 - r'^2)^{1\over2}}{r_+^2}
\cosh\frac{r_+\Delta t}{\ell^2}
+ i \epsilon {\rm ~sign~}\Delta V
\right]^{-{1\over2}}.
\nonumber
\end{eqnarray}
Of course in this region
${\rm sign~}\Delta V \neq {\rm sign~} \Delta t$.
For $x \in$ region R and  {$x' \in$ region F, we have
\begin{eqnarray}
G_1^+(x,x') &=& \sum_{n=-\infty}^\infty
\left[ \frac{r r'}{r_+^2} \cosh\frac{r_+(\Delta \phi + 2 \pi n)}{\ell}
- 1 - \frac{(r^2 - r_+^2)^{1\over2} (r_+^2 - r'^2)^{1\over2}}{r_+^2}
\sinh\frac{r_+\Delta t}{\ell^2}
+ i \epsilon {\rm ~sign~}\Delta V
\right]^{-{1\over2}} \nonumber\\
G_2^+(x,x') &=& \sum_{n=-\infty}^\infty
\left[ \frac{r r'}{r_+^2} \cosh\frac{r_+(\Delta \phi + 2 \pi n)}{\ell}
+ 1 - \frac{(r^2 - r_+^2)^{1\over2} (r_+^2 - r'^2)^{1\over2}}{r_+^2}
\sinh\frac{r_+\Delta t}{\ell^2}
+ i \epsilon {\rm ~sign~}\Delta V
\right]^{-{1\over2}}.
\nonumber
\end{eqnarray}
In all of these expressions,
$G^-(x,x')=\langle 0|\phi(x')\phi(x)|0\rangle$
is obtained by reversing the sign of $i\epsilon$.
All of these expressions are uniformly convergent for $x, x'$ real, and
$r,r'>\delta$, $\delta>0$.
Notice that as $M\to0$ ($r_+\to 0$),
$G(x,x')$ will diverge like $\sum \frac{1}{n}$
unless we take the Dirichlet boundary conditions.

{}From these expressions the Feynman Green's function can easily be
constructed and in fact has exactly the same form, but with the sign of
$i\epsilon$ being strictly positive. It should be noted that none of these
Green's functions are invariant under Anti de Sitter transformations, as
the Killing vector field defining the identification does not commute with
all the generators of the AdS group.

\subsection{KMS condition}
\label{sec:iii.2}

A thermal noise satisfies a skew periodicity in imaginary time called
Kubo-Martin-Schwinger (KMS) condition \cite{KMS}
\[
g_\beta(\Delta\tau - \frac{i}{T}) = g_\beta(-\Delta\tau)
\]
where $g_\beta(\Delta\tau) = G_\beta^{+} (x(\tau), x(\tau'))$
and $G_\beta^+ = \langle 0_\beta | \phi (x) \, \phi (x') |0_\beta \rangle
\Big|_{{\rm Im~} x^0 = -\epsilon}$ with the world line $x(\tau)$ taken
to be the one at rest with respect to the heat bath
(for a more extensive
discussion of the KMS condition, see \cite{Takagi}).
We will show that
$g(\Delta \tau) = G^+_A (x(\tau),x(\tau'))$ with $x(\tau)=
(\frac{\tau}{b},r,\phi)$
and $b=(r^2-r_+^2)^{1/2}/\ell$,
satisfies this condition outside the horizon, with a local temperature
$T= \frac{r_+}{2 \pi \ell(r^2-r_+^2)^{1/2}}$, which agrees with the Tolman
relation \cite{Tolman} $T=(g_{00})^{-1/2}T_0$, with $T_0={r_+/2\pi\ell^2}$
the temperature of the black hole.

$g(\Delta \tau)$ is defined as
\[
g(\Delta \tau) =
\frac{1}{4\sqrt{2} \pi \ell} (g_1 (\Delta \tau) \pm g_2 (\Delta \tau))
\]
where
\begin{eqnarray}
g_1(\Delta \tau) = \sum_{n=-\infty}^\infty \left[
\frac{r^2}{r_+^2} \cosh
\frac{2\pi n r_+}{\ell}
- 1 -
\frac{(r^2 - r_+^2)}{r_+^2} \cosh \left(
\frac{r_+}{\ell^2} (\frac{\Delta \tau}{b}
 - i \epsilon)\right)
\right]^{-{1\over2}} \nonumber
\\
g_2(\Delta \tau) = \sum_{n=-\infty}^\infty \left[
\frac{r^2}{r_+^2} \cosh
\frac{2\pi n r_+}{\ell}
+ 1 -
\frac{(r^2 - r_+^2)}{r_+^2} \cosh \left(
\frac{r_+}{\ell^2} (\frac{\Delta \tau}{b}
 - i \epsilon)\right)
\right]^{-{1\over2}}.
\nonumber
\end{eqnarray}
We will demonstrate the KMS
property for each term in these sums.

Take a typical term in the sum.  It has singularities at
\[
\Delta \tau_n = \pm \Delta \tau_n^0 + \frac{i}{T}p+i\epsilon
\]
where $p$ is an integer.
These singularities are square root
branch points and the branch cuts go from
$\left(\Delta \tau_n^{\,0} + 2\pi \frac{i}{T} p + i\epsilon
\to \infty + \frac{i}{T} p + i\epsilon \right)$ and
$\left(-\Delta \tau_n^{\,0} + 2\pi \frac{i}{T} p + i\epsilon
\to -\infty - 2\pi \frac{i}{T} p + i\epsilon \right)$.
In any region without the branch cuts, $g_1$ and $g_2$ are analytic. Going
around a branch point gives an additional minus sign.  Now for a given $n$, if
$\Delta \tau$ is such
that the expression inside the square root is positive,
then
$|{\rm Real~} \Delta \tau | < \Delta \tau_n$.
In this region, $g_1^n$ and
$g_2^n$ are
analytic and periodic
in $\frac{i}{T}$.  What's more $g^n(-\Delta \tau) = g^n(\Delta
\tau)$ as $\epsilon \to 0$.  If on the other hand the expression inside the
square root is negative, then because of the branch cuts,
$g^n(\Delta \tau - \frac{i}{T}) = -g^n(\Delta \tau)$.
As $g^n(\Delta \tau) = (-A + i\epsilon
{\rm ~sign~} \Delta \tau)^{-{1\over2}}$
and because our definition of the square root is with a branch cut along the
negative real axis, we see that $g^n(\Delta \tau) = -g^n (-\Delta \tau)$
($A$ is only a function of $|\Delta \tau|$).
This shows that the KMS condition is satisfied, and hence that
$G^{+}$ is a thermal
Green's function.

\subsection{Identifying the Vacuum State}
\label{sec:iii.3}

In the region R where $r>r_+$,
the Kruskal co-ordinates are defined as
\begin{eqnarray}
U &=& \left( \frac{r - r_+}{r + r_+} \right)^{\!{1\over2}}
\cosh \frac{r_+}{\ell^2} t \nonumber
\\
V &=& \left( \frac{r - r_+}{r + r_+} \right)^{\!{1\over2}}
\sinh \frac{r_+}{\ell^2} t.
\nonumber
\end{eqnarray}
Defining $\bar{V} = V - U$ and $\bar{U} = V + U$, $r$ is given by
\[
\frac{r}{r_+} = \frac{1-\bar{U}\bar{V}}{1 + \bar{U}\bar{V}}.
\]
In these co-ordinates the two point function becomes
\begin{eqnarray}
&&G_{J}^+ (\bar{U}, \bar{V}, \phi; \bar{U}' \bar{V}', \phi') =
\frac{1}{\sqrt{2}4\pi\ell}
{}~ \sum_n ~\Biggl\{
\frac{1}{(1+\bar{U}\bar{V})(1 + \bar{U}'\bar{V}')} \times \nonumber\\
&& \Biggl[
(1-\bar{U}\bar{V})
(1-\bar{U}'\bar{V}')
\cosh \left(\frac{r_+}{\ell}
(\Delta \phi + 2 \pi n)\right) \mp
(1+\bar{U}\bar{V})
(1+\bar{U}'\bar{V}')
+ 2(\bar{V}\bar{U}' + \bar{U}\bar{V}')
+ 2 i\epsilon {\rm ~sign}\,\Delta V \Biggr] \Biggr\}^{-\frac{1}{2}}
\nonumber
\end{eqnarray}
where $\mp$ is for $J=1,2$.
Here the sign of $i\epsilon$ is the same as ${\rm sign}\,\Delta V$
which is the same as ${\rm sign} \Delta \lambda$ for timelike separated points.
For $x,x' \in {\rm R}$, this is just
${\rm sign}\Delta t = {\rm sign}\,(\bar{V}\bar{U}' - \bar{U}\bar{V}')$.
This expression is valid all over the Kruskal manifold.

As discussed in the Introduction, the Hartle-Hawking
Green's function is defined to be
analytic and bounded in the lower half complex plane of $\bar{V}$ on the
past horizon $(\bar{U}=0)$, when $\bar{U}',\bar{V}',\phi,\phi'$ are real,
or in the lower half plane of $\bar{U}$ on the future horizon
($\bar{V}=0$).

On the past horizon $\bar{U}=0$ we have
\[
G^+_J = \frac{1}{\sqrt{2} 4 \pi \ell}
\sum_n
\Biggl\{
\frac{1}{1+\bar{U}'\bar{V}'}
\left[
(1-\bar{U}'\bar{V}') \cosh \left(\frac{r_+}{\ell} (\Delta \phi + 2 \pi n)
\right)\mp
(1+\bar{U}'\bar{V}') + 2 \bar{V}\bar{U'} + 2 i \epsilon \Delta V
\right]
\Biggr\}^{-{1\over2}}.
\]
In order to prove analyticity and boundedness we will show that the
singularities occur in the upper half plane of $\bar{V}$.
Hence every term in
the sum is a holomorphic function in the lower half plane. We will
then use Wierstrass's
Theorem on sums of holomorphic functions in order to prove that
the $G_J$ are analytic in the lower half of the complex $\bar{V}$ plane.

$G_J^{\,+}$ has singularities when
\[
\bar{V} =
\frac
{\pm (1+\bar{U}'\bar{V}') - (1-\bar{U}'\bar{V}')
\cosh \left(\frac{r_+}{\ell} (\Delta \phi + 2 \pi n)\right)}
{2(1+i\epsilon)\bar{U}'},
\]
Now suppose that
$x' \in {\rm R}$, then $-1 \leq \bar{U}'\bar{V}' <0$ and $\bar{U}' > 0$.
Defining
$\bar{U}\bar{V}' = -a$, $(1>a>0)$, the singularity occurs at
\[
\bar{V} = \frac
{\pm (1-a) - (1+a) \cosh
\left(\frac{r_+}{\ell} (\Delta \phi + 2 \pi n)\right)}
{2 (1 + i\epsilon) \bar{U}'}.
\]
We see that when $\epsilon \to 0$, $\bar{V}$ is real and negative.
Hence
$\bar{V} = \frac{-A}{1+i\epsilon} \simeq -A + i\epsilon$ with
$A>0$, so that the
singularities are in the upper half plane.
Similarly, for the future horizon $\bar{V}=0$, there are singularities when
\[
\bar{U} = \frac
{\pm (1+\bar{U}'\bar{V}') - (1-\bar{U}'\bar{V}') \cosh (\Delta\phi + 2\pi n)}
{(1-i\epsilon)\bar{V}'}.
\]
For $x' \in {\rm R},$ then $-1 < \bar{U}'\bar{V}' < 0,$ and $ \bar{V}'<0$,
so that $\bar{U} = \frac{A}{1-i\epsilon} = A + i\epsilon$, with $A>0$,
so the singularities are in the upper half plane of $\bar{U}$.
At this point it should be noted that for $G_J^{\,-}$ we get
singularities in the lower half plane of $\bar{U}$ on the surface
$\bar{V} = 0$, and singularities in the lower half plane of $\bar{V}$
on the surface $\bar{U}=0$.

For $x' \in {\rm F}$, if $\bar{U}=0$ and $x$ and $x'$ connected by a
null geodesic, then $\Delta V < 0$.
This is the case because for timelike and null
separations, ${\rm sign}\,\Delta V = -{\rm sign}\,\Delta r$ ($r$ is a
timelike co-ordinate in F) and $\Delta r$ is always positive if $x$ is on the
horizon.
Then it can be checked that the singularities are again
in the upper half plane of either $\bar{U}$ or $\bar{V}$.

Now that we have established that each term in the infinite sum is holomorphic
in the lower half plane of $\bar{V}$ on the past horizon (and in $\bar{U}$ on
the future horizon) we will use Weierstrass's Theorem.
This states \cite{Allfors} that if a series with analytic terms
\[
f(z) = f_1(z) + f_2(z) + \cdots
\]
converges uniformly on every compact subset of a region $\Omega$,
then the sum $f(z)$ is analytic in $\Omega$, and the series can be
differentiated term by term.
It is easily seen that unless $\bar{U'}\bar{V'}=1$, {\it i.e.} $x'$ is at
the singularity, the sum converges uniformly on every compact subset of
the lower half plane. For $\bar{U'}\bar{V'}=1$ the sum diverges and the
Green's function becomes singular at $r=0$. This is because $r=0$ is a
fixed point of the identification.

To conclude we have shown that our
Green's function is analytic on the past horizon in the lower half
$\bar{V}$ complex plane, and similarly on the future horizon in the lower half
$\bar{U}$ complex plane.
Its singularities occur when $x,x'$ can be connected by a null geodesics
either directly or after reflection at infinity (see Appendix A and
Ref.~\cite{GibbnPerr}).
We conclude that the Green's function we have constructed is
the Hartle-Hawking Green's function as defined in the Introduction, for
both Neumann and Dirichlet boundary conditions.

\subsection{The $M=0$ Green's function }
\label{sec:iii.4}

The black hole solution as $M \to 0$ is the spacetime with metric
\cite{BTZ}
\[
ds^2 =
- \left( \frac{r}{\ell} \right)^2 dt^2
+ \left( \frac{\ell}{r} \right)^2 dr^2
+ r^2 d\phi^2
\]
with $r>0$, and $t$ and $\phi$ as in (\ref{2.1}).
Defining $z=\frac{\ell}{r}$ and $\gamma = \frac{t}{\ell}$
the metric becomes
\begin{equation}
ds^2 = \frac{\ell^2}{z^2} ( - d\gamma^2 + dz^2 + d\phi^2 ).
\label{2.38}
\end{equation}

The modes for a massless conformally coupled scalar field
are solutions of the equation
\[
\Box \phi - \frac{1}{8} R \phi = 0
\]
where again $R=-6\ell^{-2}$, and are given by
\[
\phi_{k m}
= N_\omega \sqrt{\frac{z}{l}} \, e^{-i\omega t} e^{i m \phi} e^{i k z}
\]
where $\omega^2 = k^2 + m^2$,
$m$ is an integer, and
$N_\omega = (8\pi^2 \omega)^{\frac{1}{2}}$ is a normalization constant
such that
$(\psi_{m k},\,\psi_{m'k'}) = \delta_{m m'} \delta (k-k')$.

As in quantization on AdS$_3$,
care must be taken at the boundary $z=0$, which is at spatial infinity.
The metric (\ref{2.38}) is conformal to
Minkowski spacetime with one spatial co-ordinate periodic and the other
restricted to be greater than zero. As in the case of AdS$_3$, we impose the
boundary conditions
\[
\frac{1}{\sqrt{z}} \psi= 0
\qquad
{\rm or}\qquad
\frac{\partial}{\partial z} (\frac{1}{\sqrt{z}} \psi (z) ) = 0
\]
at $z=0$,
corresponding to Dirichlet or Neumann boundary conditions in the conformal
Minkowski metric.
Our approach will be to first
calculate the Green's function without boundary conditions
and them use the method of images
to impose them.

Summing modes, we obtain the two point function
\begin{eqnarray}
\tilde{G}(x,x') &=& \frac{1}{8\pi^2 \omega} ~ \frac{\sqrt{z z'}}{\ell}
{}~~\sum_m
{}~\int_k
e^{-i\omega\Delta\gamma} \, e^{i m \Delta \phi} \, e^{i k \Delta z} \, dk
\nonumber\\
&=& \frac{1}{2\pi} ~ \frac{\sqrt{z z'}}{\ell}
{}~~\sum_m
{}~e^{i m \Delta \phi}
{}~G_2 (y,y',m)
\nonumber
\end{eqnarray}
where $G_2(y,y',m)$ is the massive 1+1 dimensional
Green's function and $y=(\gamma,z)$.

Now \cite{BnD},
\begin{eqnarray}
G_2(y,y',m) &=& \frac{1}{2\pi} K_0 (|m| \, d) \qquad m\neq0 \nonumber
\\
&=& -\frac{1}{2\pi} \log d + \lim_{n\to2}
\frac{\Gamma (\frac{n}{2} -1)}{n\pi^{\frac{n}{2}}}\qquad m=0
\nonumber
\end{eqnarray}
where $d=\epsilon + i\Delta {\rm ~sign}\,\Delta t$,
with $\Delta = ( (\Delta\gamma)^2 - (\Delta z)^2 )^{\frac{1}{2}}$
for timelike separation,
and $d = \left((\Delta z)^2 - (\Delta
\gamma - i \epsilon)^2\right)^{{\frac{1}{2}}}$
for spacelike separation. Here $K_0$ is a modified Bessel function.
It follows that
\[
\tilde{G}(x,x') = \frac{1}{4\pi^2} \frac{\sqrt{z z'}}{\ell}
\left[ 2 \sum_{m>0} \cos (m \Delta \beta) K_0 (md) - \log d \right]
\]
where the infinite constant in the $m=0$ expression was dropped to
regularize the infrared divergences of the 1+1 dimensional Green's
function.
Using \cite{HandR}
\[
\sum_{m=1}^{\infty} K_0(mx) \cos (mxt) = \frac{1}{2} (c + \ln \frac{x}{4\pi})
+ \frac{\pi}{2\sqrt{x^2 + (xt)^2}}
+ \frac{\pi}{2} \sum_{l\neq0}
\left[
\left( x^2 + (2\ell\pi - tx)^2 \right)^{-\frac{1}{2}} - \frac{1}{2\ell\pi}
\right].
\]
Here $c$ is the Euler constant. This expression is valid for $x>0$
and real $t$, and gives
the following expression for spacelike separated $x$ and $x'$:
\begin{eqnarray}
\tilde{G} (x,x') &=& \frac{1}{4\pi} \frac{\sqrt{z z'}}{\ell}
\left[
\sum_n
\left[
(\Delta z)^2 + (\Delta\phi + 2\pi n)^2 - (\Delta \gamma-i\epsilon)^2
\right]^{-\frac{1}{2}}
-\sum_{n\neq0} \frac{1}{2\pi n} + c_1
\right]\nonumber\\
&=&
\frac{1}{4\pi} \frac{\sqrt{z z'}}{\ell} F(x,x')
\nonumber
\end{eqnarray}
Here $c_1=c-\ln 4\pi$.
Although the above formula was only true for $x>0$ and real $t$, the result
is analytic for every real $\Delta z$, $\Delta\gamma$ and $\Delta\phi$.
Hence it is also correct for timelike separated points.

This is just what is expected from the conformality to the Minkowski space,
other than the $\sum\frac{1}{2\pi\ell} - c_1$ which regularizes $\tilde{G}$.
Now the boundary condition can be easily put in by writing
\[
G^+(x,x')=\frac{1}{4\pi} ~ \frac{\sqrt{z z'}}{\ell} ~
\left( F(x,x') \pm F(x,\bar{x}') \right)
\]
where
$\bar{x}'=(\gamma',-z',\phi)$.
Notice that for Dirichlet boundary conditions, this agrees with the $M\to
0$ limit of (\ref{3.13}) and (\ref{3.14}).

Going to the $(t,r,\phi)$ co-ordinates we have
\[
G^+_{M=0} =\frac{1}{4\pi} (r r')^{-\frac{1}{2}} (G^+_1 \pm G^+_2)
\]
with
\begin{eqnarray}
G^+_1 &=& \sum_n \left[ \ell^2 \left(\frac{r'-r}{rr'}\right)^2
+ (\Delta\phi+2\pi n)^2
- \left(\frac{\Delta t - i \epsilon}{\ell}\right)^2
\right]^{-\frac{1}{2}} - \sum_{n\neq0} \frac{1}{2\pi n} + c_1 \nonumber
\\
G^+_2 &=& \sum_n \left[ \ell^2 \left(\frac{r'+r}{rr'}\right)^2
+ (\Delta\phi+2\pi n)^2
- \left(\frac{\Delta t - i \epsilon}{\ell}\right)^2
\right]^{-\frac{1}{2}} - \sum_{n\neq0} \frac{1}{2\pi n} + c_1
\nonumber
\end{eqnarray}

\subsection{Computation of $\langle\phi^2\rangle$}
\label{sec:iii.5}
$\langle\phi^2\rangle$ is
defined as $\langle\phi^2\rangle=\lim_{x\to x'}
{\frac{1}{2}} G_{\rm Reg}(x,x')$ where $G=G^++G^-$
is the symmetric Green's function.
In order to compute $\langle\phi^2\rangle$, we need to regularize $G$.
Now only the
$n=0$ term in $G^+_1$ is infinite and is just a Green's function on AdS$_3$.
Hence, we
can use the Hadamard development in AdS$_3$ to regularize $G$
\cite{BnD}:
\[
G_{\rm Had}=\frac{-i}{2\sqrt{2}\pi}\frac{\Delta^{\frac{1}{2}}}{
\sigma^{\frac{1}{2}}}
\]
where
\[
\sigma=\frac{\ell^2}{2}\left[ar\cos Z\right]^2,\qquad
\Delta^{-{\frac{1}{2}}}=
\frac{\sin\left(\frac{2\sigma}{\ell^2}\right)^{\frac{1}{2}}}{\left(\frac{2
\sigma}{\ell^2}\right)^{\frac{1}{2}}}\qquad{\rm and}\qquad
Z=\frac{\cos\Delta\lambda-\sin\rho\sin\rho'\cos\Delta\theta}{\cos\rho
\cos\rho'}
\]
(here $\Delta$ is the Van Veleck determinant). Defining
\[
G_{\rm Reg}(x,x')=G_{\rm BH}(x,x')-G_{\rm Had}(x,x')
\]
we get
\[
\langle\phi^2\rangle=\frac{1}{4\sqrt{2}\pi\ell}\frac{r_+}{r}\left[
\sum_{n\ne 0}\left(\cosh\left(\frac{r_+}{\ell}2\pi n
\right)-1\right)^{-{\frac{1}{2}}}\pm
\sum_{n}\left(\cosh\left(\frac{r_+}{\ell}2\pi
n\right)-1+2\left(\frac{r_+}{r}\right)^2\right)^{-{\frac{1}{2}}}\right]
\]
which, for Dirichlet boundary conditions,
can be seen to be regular as $M\to
0$ (that is $r_+\to 0$),
and to coincide in this limit with the $M=0$ result
for Dirichlet boundary conditions.

\section{The Energy-Momentum Tensor}
\label{sec:iv}

The energy-momentum tensor for a massless conformally coupled scalar field in
AdS$_3$ is given by the expression
\[
T_{\mu\nu} (x) = \frac{3}{4} \partial_\mu \phi(x) \partial_\nu \phi(x)
- \frac{1}{4} g_{\mu\nu} g^{\rho\sigma} \partial_\rho \phi(x) \partial_\sigma
\phi(x) - \frac{1}{4} \nabla_\mu \partial_\nu \phi(x) \phi(x) +
\frac{1}{96} g_{\mu\nu} R \phi^2 (x)
\]
where $R=-6\ell^{-2}$.
In order to compute $\langle T_{\mu\nu}\rangle$
one differentiates the symmetric two-point
function $G= \langle 0 | \phi(x)\,\phi(x') + \phi(x')\,\phi(x) | 0 \rangle$
\cite{BnD}, and then takes the coincident point limit.
This makes $\langle T_{\mu\nu}\rangle$
divergent and regularization is needed.  A look at our Green's
function reveals that only the $n=0$ term in $G_1$ diverges as $x \to
x'$, so only the $\langle T_{\mu\nu}\rangle$
derived from it should be regularized.

The $n=0$ term is just the Green's function in AdS$_3$ in accelerating
co-ordinates.  The vacuum in which this Green's function is derived is
symmetric
under the Anti de Sitter group and AdS$_3$ is a maximally symmetric space.
Hence
\cite{Wienberg}
$\langle T_{\mu\nu}\rangle = \frac{1}{3}
g_{\mu\nu}\langle T\rangle$ where $\langle T\rangle
=g^{\mu\nu}\langle T_{\mu\nu}\rangle$. For a
conformally coupled massless scalar field $\langle T\rangle=0$
(there is no conformal
anomaly in 2+1 dimensions)
so $\langle T_{\mu\nu}^{\rm AdS}\rangle = 0$.

Having shown that we may drop the $n=0$ term in $G_1$,
after a somewhat lengthy
calculation we arrive at the result for $M\neq 0$,
\begin{eqnarray}
\langle T_\mu^\nu(x)\rangle = \frac{1}{16\pi \ell^3r^3}
\sum_{n>0}\Biggl\{\left[\frac{r_+^2f_n^{-1}}{2}\left[
1\pm \left(1+\left(f_nr\right)^{-2}\right)^{-{3\over
2}}\right]+f_n^{-3}\right] {\rm ~diag}(1,1,-2)&&\nonumber
\\
\pm{3\over 2}\left(1-{r_+^2\over r^2}\right)f_n^{-3}
\left(1+\left(
f_nr\right)^{-2}\right)^{-{5\over 2}}{\rm ~diag}(1,0,-1)&&\Biggr\}
\label{4.2}
\end{eqnarray}
where
$f_n= \sinh (\frac{r_+}{\ell}\pi n )/r_+$.
${\rm diag}\,(a,b,c)$ is in $(t,r,\phi)$ co-ordinates. As expected
the $n=0$ term from $G_2$ did not contribute.

For $M=0$ we get from Sec.~\ref{sec:iii.4}
\begin{eqnarray}
&&\langle T^{\mu}_{\nu} (x) \rangle= \frac{1}{16\pi r^3}\sum_{n>0}
\Biggl\{
\frac{1}{(n\pi)^3}
{\rm ~diag}\,(1,1,-2)
\pm \frac{3}{2(n\pi)^3}
\left(1+\left(f_nr\right)^{-2}
\right)^{-\frac{5}{2}}
{\rm ~diag}\,(1,0,-1) \Biggr\}
\label{4.8}
\end{eqnarray}
where now $f_n=\pi n/\ell$.
Note that the $M=0$ result agrees with the $M\to 0$ limit of (\ref{4.2}).

Some properties of $\langle T_\mu^\nu\rangle$ are:
\begin{itemize}
\item As we can see from (\ref{4.2}), far away from the black hole,
$\langle T^{\mu}_\nu\rangle$
obeys the strong energy condition \cite{HandE}
only for the Dirichlet boundary
conditions, while
for the Neumann boundary conditions, the energy density
is negative in this limit.
\item For Dirichlet
boundary conditions, as $M$
decreases, although the temperature decreases, the energy density
increases; just the opposite occurs
for Neumann boundary conditions.
\item In the limit $M\to\infty$,
$\langle T^\mu_\nu\rangle\to 0$ for both sets of boundary conditions,
which suggests the presence of a Casimir effect.
\item On the horizon, $\langle T^\mu_\nu\rangle$ is
regular, and hence in the semiclassical approximation, the horizon is
stable to quantum fluctuations;
on the other hand, at $r=0$, $\langle T^\mu_\nu\rangle$ diverges.
\item Our Green's function was thermal in
$(t,r,\phi)$ co-ordinates, but although
$\langle T^\mu_\nu\rangle\sim T^3_{\rm loc}$ for large
$r$, it is not of a thermal type \cite{Tolman}.
\end{itemize}

\section{The response of a Particle detector}
\label{sec:v}

In this section we calculate the response of a particle detector which is
stationary in the black hole co-ordinates $(t,r,\phi)$, and
outside the black hole.  The
simplest particle detector can be described by an idealized point monopole
coupled to the quantum field through an interaction described by
${\cal L}_{\rm int} = c m (\tau) \, \phi[x(\tau)]$
where $\tau$ is the detector's proper time, and
$c \ll 1$.  The probability per unit time for the detector to undergo a
transition from energy $E_1$ to $E_2$ \cite{BnD} is
$R(E_1/E_2)=c^2\,\left|\langle E_2|m(0)|E_1\rangle\right|^2\,F(E_2-E_1)$
to lowest order in perturbation theory, where
\[
F(\omega) = \lim_{s\to0} ~ \lim_{\tau_0\to\infty}
{}~ \frac{1}{2\tau_0}
\int_{-\tau_0}^{\tau_0}
d\tau
{}~ \int_{-\tau_0}^{\tau_0}
d\tau'\,
e^{-i\omega(\tau-\tau')-S|\tau|-S|\tau'|}\,
{}~ g(\tau,\tau').
\]
$g(\tau,\tau') = G^{+}(x(\tau),x(\tau'))$ and $x(\tau)$
is the detector trajectory.

$F(\omega)$ is called the response function.  It represents the bath of
particles that the detector  sees during its
motion \cite{Unruh}. We take $x(\tau) = (\frac{\tau}{b},r,\phi)$
where $b = \left( \frac{r^2 - r_+^2}{\ell^2} \right)^{\frac{1}{2}}$.
Because
$g(\tau,\tau') = g(\Delta\tau)$, then
\begin{equation}
F(\omega)
= \int_{-\infty}^{\infty} e^{-i\omega\Delta\tau}
\, g(\Delta\tau) d(\Delta\tau)
\label{5.2}
\end{equation}
where $g(\Delta\tau) = g_1 (\Delta\tau) \pm g_2 (\Delta\tau)$.
Here
\begin{equation}
g_i(\Delta\tau) = \frac{r_+}{\sqrt{2} 4 \pi \ell}
(r^2 - r_+^2)^{-\frac{1}{2}} \sum_n
\left[ \frac{r_+^2}{r^2 - r_+^2}
\left( \frac{r^2}{r_+^2} \cosh \left(\frac{r_+}{\ell}2 \pi n \right)
\mp 1 \right)
- \cosh \frac{r_+}{\ell^2} (\frac{\Delta\tau}{b} - i\epsilon)
\right]^{-\frac{1}{2}}.
\label{5.3}
\end{equation}
In this expression $-(+)$ is for $g_1(g_2)$.

Defining
\[
\cosh \alpha_n = \frac{r_+^2}{r^2 - r_+^2}
\left( \frac{r^2}{r_+^2} \cosh \left(\frac{r_+}{\ell} 2\pi n\right)
 - 1 \right)
\]
and
\[
\cosh \beta_n = \frac{r_+^2}{r^2 - r_+^2}
\left( \frac{r^2}{r_+^2} \cosh\left( \frac{r_+}{\ell} 2\pi n \right)
+ 1 \right)
\]
we have from Appendix B that
\[
F(\omega) = \frac{1}{2} ~ \frac{1}{e^{\omega/T}+1} \, \sum_n
\left(
P_{\frac{i\omega}{2\pi T} - \frac{1}{2}}  ({\cosh} \alpha_n)
\pm
P_{\frac{i\omega}{2\pi T} - \frac{1}{2}}  ({\cosh}  \beta_n)
\right)
\]
where
$T=\frac{r_+}{2\pi\ell (r^2 - r_+^2)^{\frac{1}{2}}}$
is the local temperature.
This looks like a fermion distribution with zero chemical potential
and a density of
states
\[
D(\omega) = \frac{\omega}{2\pi} ~ \sum_n
\left(
P_{\frac{i\omega}{2\pi T} - \frac{1}{2}}  ({\cosh} \alpha_n)
\pm
P_{\frac{i\omega}{2\pi T} - \frac{1}{2}}  ({\cosh}  \beta_n)
\right).
\]
Notice that
$F(\omega)$ is finite on the horizon in contrast with black holes in two
and four dimensions (see \cite{candelas}). This seems to be a consequence
of the Fermi type distribution.
Statistical inversion in odd-dimensional flat spacetime
was first noted in Ref.~\cite{Takagi}.

If the mass of the black hole satisfies $e^{2\pi\sqrt{M}}\gg 1$ and
$2\ell^{-1}>\omega\gg T$,
then far from the horizon, $r \gg r_+$, we can sum the series, and
for Dirichlet boundary conditions we obtain
\[
F(\omega) \simeq 2\pi^2\ell^2T^2\frac{1}{e^{\omega/T}+1}
\left[
\left( \frac{\omega}{2\pi T} \right)^2 + \frac{1}{4}
+ \frac{8e^{-3\pi\sqrt{M}}}{\pi}
\left( \frac{\omega}{T} \right)^{\frac{1}{2}}
\left( \sin \frac{w\sqrt{M}}{T} - \cos \frac{w\sqrt{M}}{T} \right)
\right].
\]
A similar result holds
for Neumann boundary conditions at large $r$,
\[
F(\omega) \simeq \frac{1}{e^{\omega/T} +1}
\left[
1+ 4\left( \frac{\omega}{T} \right)^{-\frac{1}{2}}
e^{-\pi\sqrt{M}}
\left(\sin\frac{\omega \sqrt{M}}{T}+\cos\frac{\omega\sqrt{M}}{T}\right)
\right]
\]
where the approximation improves for large $M$ as before.

It seems clear that the particle detector response will consist of a
Rindler-type effect \cite{BnD}, and, if present, a response due to
Hawking radiation (real particles). The former is due to the fact that a
stationary particle detector is actually accelerating, even
when $r\to\infty$
(there is no asymptotically flat region). This is reflected for instance
in the fact that for some range of
$\omega$, the behaviour of
$F(\omega)$ for $r\gg r_+$ and $M\gg 1$ is governed
by the $n=0$ term in $G^+$, which is AdS invariant.
Hence all observers connected by an AdS transformation (a subgroup of
the asymptotic symmetry group) register the same
response, even though they might be in relative motion; this means that
$F(\omega)$ as a whole cannot be interpreted as real particles (see
\cite{GandH,haij} for a discussion of this point). Unfortunately, one
cannot filter out these effects in a simple way, and further work is needed
in order to find the spectrum of the Hawking radiation.

Finally, for $M=0$, we may again define
\[
F_i(\omega)
= \int_{-\infty}^\infty e^{-i\omega \Delta \tau} g_i (\Delta \tau)
d \Delta \tau.
\]
Now, however,
$g_1$ and $g_2$ are analytic in the lower half complex
plane of $\Delta \tau$. Hence for $\omega>0$ we can close the integral in
an infinite semicircle in the lower half plane and by Cauchy's theorem
$F_i(\omega) = 0$ for $\omega>0$ so that no particles are detected by a
stationary particle detector.

\section{Back-reaction}
\label{sec:vi}

In this section we shall discuss the back-reaction on the BTZ solutions due
to quantum fluctuations, using the energy momentum tensor $\langle
T^\mu_\nu\rangle$ derived in Sec. \ref{sec:iv}. We shall show that for all
$M$, including $M=0$, divergences in the energy momentum tensor cause the
curvature scalar $R_{\mu\nu}R^{\mu\nu}$ to blow up at $r=0$ (note that
since the energy momentum tensor is traceless, $R$ does not blow up). It is
also interesting to consider the effects of back-reaction on the location
of the horizon, even thought this is only an order $\hbar$ effect. It is
possible to show for all $M\ne 0$ that the horizon shifts outwards under
the effect of quantum fluctuations. For $M=0$, the effect is
that a horizon develops at a radius of order $\hbar$, but where we may
still be able to trust the semi-classical approximation.

We compute the back-reaction in the usual way by inserting
the expectation value of the
energy-momentum tensor (\ref{4.2}) or (\ref{4.8}), into Einstein's equations,
\[
G_{\mu\nu}=\ell^{-2} g_{\mu\nu}+\pi\langle T_{\mu\nu}\rangle.
\]
The first thing to note is that although
the external solution is of constant curvature everywhere, the
perturbed solution is not, and the curvature scalar
$R_{\mu\nu} R^{\mu\nu}$ diverges at the origin, $r=0$.
Einstein's equations give
\begin{eqnarray}
R_{\mu\nu} R^{\mu\nu} &=&
(\pi\langle T^\mu_{\,\nu}\rangle - 2\ell^{-2}\delta^\mu_{\,\nu})
(\pi\langle T_\mu^{\,\nu}\rangle - 2\ell^{-2}\delta_\mu^{\,\nu}) \nonumber
\\
&=& \pi^2\langle T^\mu_{\,\nu}\rangle\langle T^\nu_{\,\mu}\rangle
+ 12\ell^{-4} \nonumber
\\
&>& \pi^2\langle T^r_{\,r}\rangle^2 + 12\ell^{-4}
= 12\ell^{-4} +
\frac{1}{64\ell^6r^6}
\left(\sum_{n>0}f_n^{-3}\right)^2
\nonumber
\end{eqnarray}
The sum in the last expression is a constant depending only on $M$.
In the limit as $M\to 0$, the curvature still diverges as $1/r^6$.
Although the divergence in the curvature scalar occurs precisely where the
semi-classical approximation is unreliable, the result does say that we
must go beyond semi-classical physics in order to describe the region near
$r=0$. This seems to be the natural notion of a singularity at the
semi-classical level.

Having shown that the back-reacted metric becomes singular, it remains to
look at horizons.  We
begin with a general static, spherically symmetric metric, which we take to
be
\[
ds^2 = -N^2 \, dt^2 \, + \frac{dr^2}{N^2} + e^{2A} \, d\phi^2 ~~,
\]
where $N$ and $A$ are functions of $r$ only.  A linear combination of
Einstein's equations implies that
\begin{equation}
(N^2)'' = 2\ell^{-2} + 2\pi \, \langle T^\phi_{\,\phi}\rangle.
\label{v:1}
\end{equation}
Integrating Eq.~(\ref{v:1}) once, and inserting (\ref{4.2}),
we obtain the result
\begin{equation}
(N^2)' = \frac{2r}{\ell^2} + \frac{1}{8\ell^3r^2}
\sum_{n>0}
\left\{\frac{r_+^2f_n^{-1}}{2}\left[1\pm \left(1+\left(
f_nr\right)^{-2}\right)^{-{3\over 2}}\right]+f_n^{-3}
\left[1\pm\frac{f_n^2r^2}{2}\left(1-\left(1+\left(f_nr
\right)^{-2}\right)^{-{3\over2}}\right)\right]\right\}
\label{v:2}
\end{equation}
where an integration constant has been included to make the result finite.
A second integration gives
\begin{eqnarray}
N^2 &=& \frac{r^2}{\ell^2} - {M}-\frac{1}{8\ell^3r}
\sum_{n>0}
\left\{ \frac{r^2_+f_n^{-1}}{2}
\left[1\pm\left(1+\left(f_nr\right)^{-2}\right)^{-{1\over 2}}\right]\right.
\nonumber
\\
&&
\left.+f_n^{-3}\left[1\pm{1\over 2}\left(
f_n^2r^2
\left[\left(1+\left(f_nr\right)^{-2}\right)^{1\over
2}-1\right]+\left(1+\left(f_nr\right)^{-2}\right)^{-{1\over
2}}\right)\right]\right\}
\label{v:3}
\end{eqnarray}
where the second integration constant has been set to ${M}$, and is the ADM
mass of the solution \cite{BTZ}. The two
integration constants ensure that
$N^2 \to r^2/\ell^2 -{M} + o (\frac{1}{r})$ as $r \to \infty$.

Having obtained an expression for $N$, it is also necessary to look at the
$g_{\phi\phi}$ component given by $A$.
$A$ is given in terms of $N$ by the equation
\[
A' = \frac{16\ell r^3+
\sum_{n>0}\left[\frac{r_+^2f_n^{-1}}{2}\left[1\pm\left(1+\left(
f_nr\right)^{-2}\right)^{-{3\over 2}}\right]+f_n^{-3}
\right]}{8\ell^3r^3(N^2)'}
\]
which we shall not attempt to integrate, although it is easy to see that as
$r\to \infty$, $A\to\ln r$.  The important thing to notice is
that $A'$ diverges only at $r=0$ or where $(N^2)'=0$.
If the singularity at $r=0$ is to
be taken seriously, it is important that $(N^2)'$ should not vanish for any
finite, non-zero $r$.  To see that this is indeed the case, note that
since the quantity inside the curly
brackets of Eq.~(\ref{v:2}),
is positive for all $r>0$, then so is $(N^2)'$.

Having checked that the backreaction does not cause a qualitative change
in $g_{\phi\phi}$, and having found the exact change in $N$, we may
examine the horizon structure of the new solutions.
Note that each term in the sum in (\ref{v:3})
is strictly positive, and behaves as $1/r$ at infinity and near the
origin, for any $M$. Hence, the horizon of the $M\ne 0$ solutions is
pushed out by quantum fluctuations, as compared with the
classical solution of the same ADM mass.

The $M=0$ solution, which acquires
a curvature singularity due to backreaction, also develops
a horizon.
We regard this result as being indicative of the fact that the $M=0$
solution is unstable, in the sense that the qualitative features of the
solution are changed by quantum fluctuations.
Recall that $\langle T_\mu^{\,\nu}\rangle$
in this case appears to be just the Casimir
energy of the spacetime as it is associated with a zero temperature Green's
function. The appearance of a horizon may be contrasted in an obvious way
with 4-dimensional Minkowski spacetime, regarded as the $M=0$ limit of the
Schwarzschild solution.  Minkowski spacetime has no Casimir energy
associated with it, and is stable in the above sense.

Note that as $M\to 0$, the horizon
is located in a region sufficiently close to
$r=0$ that the semi-classical approximation may break down, {\it i.e.}
fluctuations in $\langle T^\mu_\nu\rangle$
will be of the order of $\langle T^\mu_\nu\rangle$. However,
if there are $n$ independent scalar fields present, then the ratio of the
fluctuations to $\langle T^\mu_\nu\rangle$ becomes negligible
in the vicinity of the horizon, as $n$ becomes large. The
size of the perturbation on the metric near where the horizon develops may
also be estimated. It is approximately an order of magnitude smaller than the
curvature of the original solution, a result which is independent of $n$.

We end with a speculation about the endpoint of evaporation.
Notice that although classically there is a clear but puzzling
distinction between the black hole solutions
of BTZ, with $M\ge 0$, the solutions with conical singularities
of Deser and Jackiw \cite{DJ}
corresponding to $-1<M<0$, and AdS$_3$ ($M=-1$), semiclassically the
difference between the small $M$ and negative $M$ solutions is not so
marked. Our results for $M=0$ are qualitatively similar to
those of Refs. \cite{har},
where it is shown that quantum fluctuations on a conical spacetime generate
a singularity at the apex of the cone, shielded by an order $\hbar$ horizon.
One might speculate from this similarity that evaporation could
continue beyond the $M=0$ solution, perhaps ending at AdS$_3$.

\section{Conclusions}
\label{sec:vii}
In this paper we presented some aspects of quantization on the 2+1
dimensional black hole geometry. We obtained an exact expression for the
Green's function in the Hartle-Hawking vacuum and for the expectation value
of the energy-momentum tensor, but we found some difficulty in interpreting
the particle detector response as Hawking radiation. We feel that further
investigation on this question is required. If the black hole
evaporates, the results of section~\ref{sec:vi} suggest the possibility
that due to quantum fluctuations, the endpoint of evaporation
may not look like the classical $M=0$ solution.

\acknowledgements
We thank Roger Brooks and Samir Mathur for much help and encouragement.
We also acknowledge useful conversations with Mike Crescimanno,
Gary Gibbons, Esko Keski-Vakkuri and Alan Steif.

\section*{Note added}
After this work was completed, we received three related preprints:
K. Shiraishi and T. Maki, Akita Junior College preprint
AJC-HEP-18, and A. R. Steif, Cambridge preprint DAMTP93/R20,
hep-th/9308032, in which a
Green's function on the BTZ black hole spacetime is computed without
the use of boundary conditions at infinity; and
K. Shiraishi and T. Maki, Akita Junior College preprint AJC-HEP-19.

\appendix
\section{Scalar field quantization on AdS$_3$}
\label{sec:app.a}

The derivation of a scalar field
propagator on AdS$_3$ is reviewed. This computation
is complicated by the fact that AdS$_3$ is not globally hyperbolic.
In the AdS co-ordinate system defined in Sec.~\ref{sec:ii},
spatial infinity is the $\rho=\frac{\pi}{2}$ surface
which is seen to be timelike (see Fig.~\ref{fig:penroseads}).
Information can escape or leak in through
this surface in a finite co-ordinate time, spoiling the composition law
property of the propagator. In order to resolve this problem and define a
good quantization scheme on AdS$_3$,
we follow \cite{Isham} and use the
fact that AdS$_3$ is conformal to half of the Einstein Static
Universe (ESU) $R\times S^2$.

The metric of ESU is
\[
ds^2 = -d\lambda^2 + d\rho^2 + \sin^2\rho\,d\theta^2
\]
where $-\infty<\lambda<\infty$, $0<\rho\leq\pi$, and
$0 < \theta\leq 2\pi$.
Positive frequency modes on ESU are solutions of
\[
\Box \psi^{\rm E} - \frac{1}{8} R \psi^{\rm E} = 0
\]
where $R=2$, and are given by
\begin{equation}
\psi^{\rm E}_{\ell m} =
N_{\ell m} \, e^{-i\omega\lambda} \, Y^{\ell}_{m} (\rho,\theta)
\qquad\omega > 0
\label{7.3}
\end{equation}
where $Y_\ell^m$ are the spherical harmonics,
$\omega = \ell + \frac{1}{2}$, $m$ and $\ell$ are
integers with $\ell \geq 0$,~ $|m|\leq \ell$,
 and $N_{\ell m} = \frac{1}{\sqrt{2\ell+1}}$.
These modes are orthonormal in the inner product \cite{BnD}
\[
(\psi_1, \psi_2) = -i\int_\Sigma \psi_1
\raise1.5ex\hbox{$\leftrightarrow$}\mkern-16.5mu\partial_\mu \psi_2^*
[-g_\Sigma(x)]^{1\over2}
d\Sigma^\mu
\]
where $\Sigma$ is a spacelike Cauchy surface.
{\it i.e.} $(\psi_{\ell m}, \psi_{\ell' m'}) = \delta_{\ell\ell'}
\delta_{mm'}$,
$(\psi_{\ell m}, \psi_{\ell' m}^{\,*}) = 0$, and
$(\psi_{\ell m}^{\,*}, \psi_{\ell' m'}^{\,*})
= -\delta_{\ell\ell} \delta_{mm'}$.
As usual the field operator is expanded in these modes
$\phi=\sum_{\ell, m} \psi_{\ell m} a_{\ell m} +
\psi_{\ell m} a_{\ell m}^{\,+}$
so that $a,a^{+}$ destroy and create particles, and define
the vacuum state $|0\rangle_E$.

The two point function is defined as
\[
G^{+}_E (x,x') =
\,_E\langle 0 | \phi(x) \, \phi(x') | 0 \rangle_E
= \sum \, \psi^E_{\ell m} (x) {\psi^{E\, *}_{\ell m}}(x').
\]
Inserting (\ref{7.3}),
\[
G_{\rm E}^{+} (x,x')
= \sum_{\ell} \frac{1}{2\ell+1}
e^{-i(\ell+\frac{1}{2})(\lambda-\lambda')}
\sum_m ~ Y_m^\ell (\rho,\theta) ~ Y_m^{* \ell} (\rho',\theta').
\]
Using $(Y_m^\ell)^{*} = (-1)^m Y_{-m}^{\ell}$
and $\sum_{m=-\ell}^{\ell} (-1)^m Y_m^\ell (x) Y_{-m}^\ell(x')
= \frac{2\ell+1}{4\pi} P_\ell (\cos \alpha)$
where $\alpha$ is the angle between $(\rho,\theta)$ and $(\rho',\theta')$,
we get
\[
G_{\rm E}^{+} (x,x') = \frac{1}{4\pi} e^{-\frac{1}{2} (\lambda-\lambda')}
\sum_{\ell=0} e^{-i \ell (\lambda - \lambda')} P_\ell (\cos \alpha).
\]
Further, using
$\sum_{n=0}^\infty P_n (x) \, z^n = (1-2xz + z^2)^{-\frac{1}{2}}$ for
$-1<x<1$ and $|z|<1$
and as usual giving $\Delta\lambda$ a small negative imaginary part
for convergence, we get
\[
G_{\rm E}^{+} = \frac{1}{4\sqrt{2}\pi}
\left(
\cos(\Delta\lambda - i\epsilon)
- \cos\rho \cos\rho'
- \sin\rho \sin\rho'
\cos \Delta\theta
\right)^{-\frac{1}{2}}
\]
where the square root is defined with a branch cut along the negative real axis
and the argument function is between $(-\pi,\pi)$ \cite{hobson}.
{}From now we shall call this two point function
$G^{+}_{1,{\rm E}}$ and define
$G^{+}_{2,{\rm E}}(x,x') =  G^{+}_{1,{\rm E}} (\tilde{x},x')$
where $\tilde{x}=(\lambda,\pi-\rho, \theta)$. Then,
\[
G_{2,{\rm E}}^{+}
= \frac{1}{4\sqrt{2}\pi}
\left(
\cos(\Delta\lambda - i\epsilon)
+ \cos\rho \cos\rho'
- \sin\rho \sin\rho'
\cos \Delta\theta
\right)^{-\frac{1}{2}}
\]
and
$G_{2,{\rm E}}^{+}$
satisfies also the homogeneous equation
$(\Box - \frac{1}{8} R) G = 0$.
Conformally mapping these solutions to AdS$_3$, where
$G_{{\rm A}}^{+} = \sqrt{\cos\rho \cos\rho'} G_{\rm E}^{+}$
we get
\[
G_{1,{\rm A}}^{+} (x,x')
= \frac{1}{4\sqrt{2}\pi\ell}
\left(
\cos(\Delta\lambda - i\epsilon)
\sec\rho \sec\rho'
-1-
\tan\rho \tan\rho'
\cos \Delta\theta
\right)^{-\frac{1}{2}}
\]
and
\[
G_{2,{\rm A}}^{+} (x,x')
= \frac{1}{4\sqrt{2}\pi\ell}
\left(
\cos(\Delta\lambda - i\epsilon)
\sec\rho \sec\rho'
+1-
\tan\rho \tan\rho'
\cos \Delta\theta
\right)^{-\frac{1}{2}}.
\]
It can be seen that
$G_{1,{\rm A}}^{+}$ and $G_{2,{\rm A}}^{+}$
are functions of
$\sigma(x,x') = \frac{1}{2} [ (u-u')^2 + (v-v')^2 + (x-x')^2 + (y-y')^2]$,
which is the distance between the spacetime points $x,x'$ in the
4-dimensional embedding space.

In order to deal with the problem of global hyperbolicity, it was shown in
\cite{Isham} that
imposing boundary conditions on the ESU modes gives a good quantization
scheme on the half of ESU with $\rho\leq\frac{\pi}{2}$, thus
inducing a good
quantization scheme on AdS$_4$. It may be checked that this method also
works in 2+1 dimensions.
The boundary
conditions on the ESU modes are either Dirichlet
\[
\psi^{\rm E}_{\ell,m} \left(\rho=\frac{\pi}{2}\right) =0
\qquad{\rm obeyed~by~} \psi_{\ell,m}{\rm~with~}\ell+m={\rm ~odd}
\]
or Neumann
\[
\frac{\partial}{\partial\rho}\,
\psi^{\rm E}_{\ell,m} \left(\rho=\frac{\pi}{2}\right) =0
\qquad{\rm obeyed~by~} \psi_{\ell,m}{\rm~with~}\ell+m={\rm ~even}.
\]

It is easily verified that the combination
$
G^{+}_{{\rm E}} =
G^{+}_{1,{\rm E}} \pm
G^{+}_{2,{\rm E}}
$
has the right boundary condition where the
$+(-)$ signs are for Neumann (Dirichlet) boundary   conditions.

Some remarks are in order:  if $x,x'$ are restricted such that
$-\pi < \lambda(x) - \lambda(x') < \pi$ then

\begin{itemize}
\item[(1)~]
${{G_1^+}}_E$ is real for spacelike points, imaginary for timelike points
and singular for $x,x'$ which can be connected by a null geodesic.

\item[(2)~]
${G_2^+}_E$ has the same property when $x\to\tilde{x}$,
and if $0 \le\rho(x'), \rho(x) < \frac{\pi}{2}$ then $G_2$ has singularities
when $x,x'$ can be connected by a null geodesic bouncing off
$\rho=\frac{\pi}{2}$ boundary.
\end{itemize}

{}From this we see that if we take the modes in AdS$_3$ as
\[
\psi^{\rm A}_{\ell,m} =
(\cos \rho)^{\frac{1}{2}} \, e^{-i(\ell+\frac{1}{2}) \lambda}
\, Y_\ell^m (\rho,\theta)
\qquad\ell+m = {\rm ~odd~~or~~} \ell+m = {\rm ~even}
\]
then these modes give rise to a
well-behaved propagator \cite{Isham}.
The two point function is then
\[
G_{\rm A}^{+} = \sqrt{\cos\rho\cos\rho'}
(G^{+}_{1,{\rm A}} \pm G^{+}_{2,{\rm A}})
\]
where $+(-)$ are for Neumann (Dirichlet).
The two point function has singularities whenever $x,x'$ can be connected by a
null geodesic directly or by a null geodesic bouncing off infinity (null
geodesics remain null geodesics by a conformal transformation).  All other
properties listed before also stay the same.

Note that it is possible to define a quantization scheme on AdS$_3$ without
using boundary conditions ({\it i.e.} just using $G^+_{1,A}$), which is
referred to
as transparent boundary conditions in Ref. \cite{Isham}. However this
requires the use of a two-time Cauchy surface, and its physical
interpretation is unclear.

\section{Calculating the response function}
\label{sec:app.b}

We are interested in an integral of the type
\[
J(\omega)=
\frac{\ell^2 b}{r_+} \int_{-\infty}^{\infty} \,
e^{-{\frac{i\omega}{2\pi T}}t} \,
(\cosh \alpha_n - \cosh(t-i\epsilon) )^{-\frac{1}{2}} dt
\]
where $T=\frac{r_+}{2\pi\ell (r^2 - r_+^2)^{\frac{1}{2}}}$
is the local temperature.
$J(\omega)=I_1 (\omega) + I_2(\omega) + I_3(\omega)$ where $I_1$ is the
integral from $-\infty$ to $-\alpha_n$,
$I_2$ is from $-\alpha_n$ to $\alpha_n$, and
$I_3$ is from $\alpha_n$ to $\infty$.
Recall that
the square root is defined with the cut along the negative real axis. Then
\begin{eqnarray}
I_1 &=& \frac{\ell^2 b}{-i r_+} \int_{-\infty}^{-\alpha_n}
e^{-\frac{i\omega t}{2\pi T}} ~(\cosh t - \cosh \alpha_n)^{-\frac{1}{2}}
\nonumber
\\
I_3 &=& \frac{\ell^2 b}{i r_+} \int_{\alpha_n}^{\infty}
e^{-\frac{i\omega t}{2\pi T}} ~(\cosh t - \cosh \alpha_n)^{-\frac{1}{2}}
\nonumber
\\
I_2 &=& \frac{2\ell^2 b}{r_+} \int_0^{\alpha_n} \cos\frac{\omega t}{2\pi T}
{}~(\cosh \alpha_n - \cosh t)^{-\frac{1}{2}} dt.
\nonumber
\end{eqnarray}
Using \cite{HandR}
\begin{eqnarray}
\int_\alpha^\infty
\frac
{e^{-(\nu + \frac{1}{2}) t}}
{({\cosh} t-{\cosh} \alpha)^{-\frac{1}{2}}}
&=& \sqrt{2} \, Q_\nu ({\cosh} \alpha) \qquad {\rm Re}\,\nu>-1~~~~~\alpha>0
\nonumber
\\
\int_0^\alpha \frac{{\cosh} (\nu + \frac{1}{2}) t}
{({\cosh}\alpha - {\cosh}t)^{\frac{1}{2}}}
&=& \frac{\pi}{\sqrt{2}} P_\nu ({\cosh} \alpha)\qquad\alpha>0
\nonumber
\end{eqnarray}
where $P_\nu$ and $Q_\nu$ are associated Legendre functions of the first
and second kind respectively,
we get
\begin{eqnarray}
I_3 &=& -\frac{i\sqrt{2}\ell^2 b}{r_+}
Q_{\frac{i\omega}{2\pi T} - \frac{1}{2}} ({\cosh} \alpha_n)
\nonumber
\\
I_2 &=& \frac{\sqrt{2}\pi \ell^2 b}{r_+}
P_{\frac{i\omega}{2\pi T} - \frac{1}{2}} ({\cosh} \alpha_n) \nonumber
\\
I_1 &=& \frac{i\sqrt{2}\ell^2 b}{r_+}
Q_{\frac{-i\omega}{2\pi T} - \frac{1}{2}} ({\cosh} \alpha_n).
\nonumber
\end{eqnarray}
Now using $Q_\nu(z)-Q_{-\nu-1}(z)=\pi \cot (\nu\pi) P_\nu(z)$ \cite{HandR}
\[
J(\omega) =\frac{2\sqrt{2}\pi\ell^2 b }{r_+} \,
P_{\frac{i\omega}{2\pi T} - \frac{1}{2}}  ({\cosh} \alpha_n)
\frac{1}{e^{\omega/T}+1} ,
\]
and $F_{1,2}(\omega)$, defined by (\ref{5.2}) and (\ref{5.3})
in an obvious
way, are given by
\begin{eqnarray}
F_1(\omega) &=& \frac{1}{2} ~
\frac{1}{e^{\omega/T}+1} ~ \sum_{n\ne 0}
P_{\frac{i\omega}{2\pi T} - \frac{1}{2}}  ({\cosh} \alpha_n) \nonumber\\
F_2(\omega) &=& \frac{1}{2} ~
\frac{1}{e^{\omega/T}+1} ~ \sum_n
P_{\frac{i\omega}{2\pi T} - \frac{1}{2}}  ({\cosh} \beta_n). \nonumber
\end{eqnarray}

Notice that although the formulae that we used were not correct when
$\alpha=0$, nevertheless the $\alpha_0$ term came out correctly, since
\begin{eqnarray}
&&\int_{-\infty}^{\infty} e^{-i\omega t}
(1-\cosh(2\pi T t - i\epsilon))^{-\frac{1}{2}} \nonumber\\
&&=
i \int_{-\infty}^{0} e^{-i\omega t}
\left| \sqrt{2} \sinh (\pi T t - i \epsilon) \right|^{-1}
-i \int_0^\infty e^{-i\omega t}
\left| \sqrt{2} \sinh (\pi T t - i \epsilon) \right|^{-1} \nonumber\\
&&=
\int_{-\infty}^\infty e^{-i\omega t}
\left( \sqrt{2} i \sinh (\pi T t) + \epsilon \right)^{-1}. \nonumber
\end{eqnarray}
This gives \cite{Takagi}
\[
F_1^0(\omega)=\frac{T}{2}
\int_{-\infty}^{\infty}
e^{-i\omega t} ~ (2 i \sinh (\pi T t) + \epsilon)^{-1} =
\frac{1}{2} ~ \frac{1}{e^{\omega/T}+1}
\]
which is exactly what we got before as $P_\nu(1)=1$.

Combining the results for $F_1$ and $F_2$, we have
\[
F(\omega)={\frac{1}{2}}\frac{1}{e^{\omega/T}+1}\sum_n\left(
P_{\frac{i\omega}{2\pi
T}-\frac{1}{2}}(\cosh\alpha_n)\pm P_{\frac{i\omega}{2\pi
T}-\frac{1}{2}}(\cosh\beta_n)\right).
\]

\begin{figure}
\caption{A Penrose diagram of (a) the $M\ne 0$ black hole, and (b) the $M=0$
solution. Information can leak through spatial infinity, unless we impose
boundary conditions at $r=\infty$.}
\label{fig:penrosebh}
\end{figure}

\begin{figure}
\caption{A Penrose diagram of AdS$_3$. Information can leak in or out through
spatial infinity, and thus $\Sigma$ is not a Cauchy surface unless we
impose boundary conditions at $r=\infty$.}
\label{fig:penroseads}
\end{figure}

\end{document}